\documentclass[manuscript,manuscript]{acmart}
\acmSubmissionID{5015}

\AtBeginDocument{%
  }

\setcopyright{acmcopyright}
\copyrightyear{2024}
\acmYear{2024}
\acmDOI{10.1145/3613904.3642480}

\acmConference[CHI'24]{Make sure to enter the correct
  conference title from your rights confirmation email}{May 11--16,
  2024}{Honolulu, USA}
\acmPrice{15.00}
\acmISBN{978-1-4503-XXXX-X/18/06}

\definecolor{flatblue}{RGB}{52, 152, 219}

\newcommand{\systemname}{\texorpdfstring{AnalogyMate}{AnalogyMate}}

\newcommand{\szd}[1]{{\texorpdfstring{\color{black} #1}{#1}}}
\newcommand{\rv}[1]{{\texorpdfstring{\color{black} #1}{#1}}}

\begin{document}

\title[Creating Analogies with Generative AI]{Beyond Numbers: Creating Analogies to \rv{Enhance} Data Comprehension and Communication with Generative AI}


\author{Qing Chen}
\affiliation{%
  \institution{Intelligent Big Data Visualization Lab, Tongji University}
  \city{Shanghai}
  \country{China}
}
\email{qingchen@tongji.edu.cn}

\author{Wei Shuai}
\affiliation{%
  \institution{Intelligent Big Data Visualization Lab, Tongji University}
  \city{Shanghai}
  \country{China}
}
\email{shuaiwei@tongji.edu.cn}

\author{Jiyao Zhang}
\affiliation{%
  \institution{Intelligent Big Data Visualization Lab, Tongji University}
  \city{Shanghai}
  \country{China}
}
\email{zhangjiyao@tongji.edu.cn}

\author{Zhida Sun}
\affiliation{%
  \institution{College of Computer Science \& Software Engineering, Shenzhen University}
  \city{Shenzhen}
  \country{China}
}
\email{zhida.sun@outlook.com}

\author{Nan Cao}
\affiliation{%
  \institution{Intelligent Big Data Visualization Lab, Tongji College of Design and Innovation}
  \city{Shanghai}
  \country{China}
}
\email{nan.cao@gmail.com}


\renewcommand{\shortauthors}{Chen et al.}

\begin{abstract}
Unfamiliar measurements usually hinder readers from grasping the scale of the numerical data, understanding the content, and feeling engaged with the context. \rv{To enhance data comprehension and communication, we leverage analogies to bridge the gap between abstract data and familiar measurements. In this work, we first conduct semi-structured interviews with design experts to identify design problems and summarize design considerations. Then, we collect an analogy dataset of 138 cases from various online sources. Based on the collected dataset, we characterize a design space for creating data analogies.} Next, we build a prototype system, AnalogyMate, that automatically suggests data analogies, their corresponding design solutions, and generated visual representations powered by generative AI. The study results show the usefulness of AnalogyMate in aiding the creation process of data analogies and the effectiveness of data analogy in enhancing data comprehension and communication. 
\end{abstract}

\begin{teaserfigure}
  \centering
  \includegraphics[width=\linewidth]{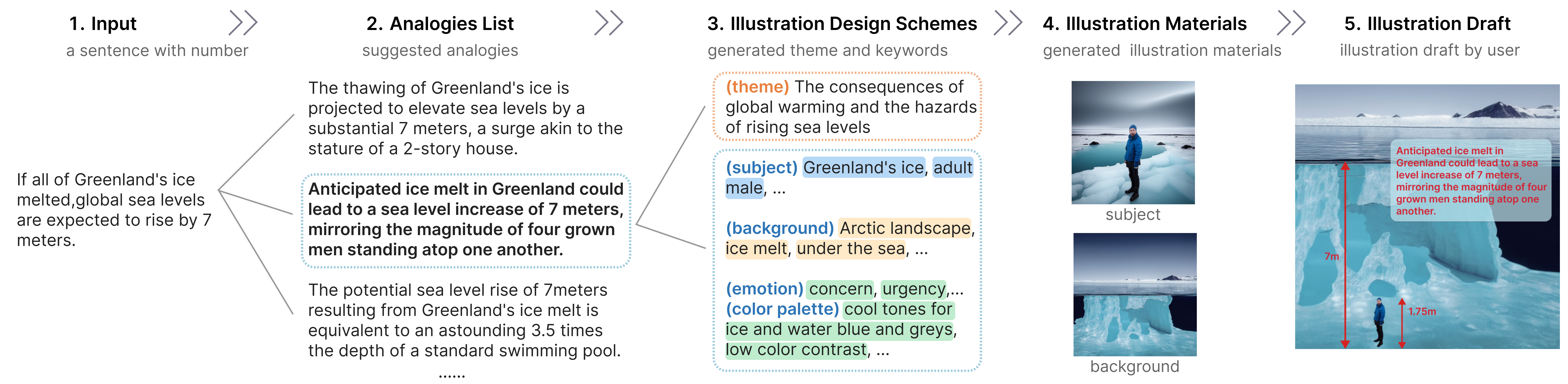}
  \vspace{-6mm}
  \caption{An example of the {\systemname} system automatically suggests data analogy for the inputs. The system first suggests data analogies according to the input, then generates corresponding design solutions and visual representations for illustrations. Finally, we show a user-rendered version of one of the analogy designs. }
  \label{fig:teaser}
\end{teaserfigure}


\maketitle
\section{Introduction}

In our daily lives, we constantly encounter a wide range of numerical data~\cite{feigenson2004core}, whether in news articles, government reports, financial statements, or scientific journals. However, numerical data's sheer scale or intricacy, often coupled with unfamiliar measurements and abstract representations, poses barriers to comprehension and communication. 
This barrier can lead to misunderstandings or misinterpretations, hindering readers from grasping the key information that authors intend to convey through the data. 
To illustrate this, consider the headline ``Every day, 1.3 billion plastic bottles are sold worldwide, stacking up to the height of half the Eiffel Tower'', with an illustrated figure shown in Fig.~\ref{fig:DataAnalogyExample}(a). The primary purpose of this example is to enable the readers to perceive the numerical data and realize the underlying significance of the environmental damage caused by such a large amount of plastic bottles. By introducing such a data analogy and its corresponding visual presentation, we bridge the gap between the abstract data ``1.3 million'' and the familiar measurements ``the height of half the Eiffel Tower''. In doing so, readers could better grasp the scale of the numerical data, enhancing their understanding of the content while feeling more engaged with the context.

The data analogy technique, involving juxtaposing abstract data with familiar objects or measurements, has become a powerful tool to improve the effective communication of data significance. This approach is rooted in understanding and empirical research that analogies serve as cognitive bridges, facilitating the comprehension and communication of numerical information~\cite{riederer2018put}. 
Despite the benefits of data analogies, it remains challenging for amateurs to design and create them with infographics. The creation process involves two stages. 
First, the ideation stage requires designers to develop appropriate analogical objects with proper measurements and scales, demonstrating the associations between the original object and the numerical data. 
Second, the design stage demands diverse visual design skills and knowledge from designers to produce visual representations with proper layout and contrast to convey the message. Designers used to search for relevant ideas, numbers, and images on multiple online resources and build everything from scratch. 
This process is usually time-consuming for experts and even more challenging for amateurs. 

Previous work has proposed some automatic methods to improve efficiency in the analogy design process. 
For example, Kim et al. \cite{kim2016generating} developed tools for creating personalized spatial analogies using a user's location and a comprehensive landmark dataset to provide contextual information for spatial measurements. Hullman et al. \cite{hullman2018improving} introduced a set of tools designed to automatically generate re-expressions for unfamiliar measurements by leveraging measurements of familiar objects. 
However, existing approaches have often depended on compiling and curating object-measurement databases within specific domains, limiting their capacity to generate diverse and innovative design solutions. Furthermore, there is a significant demand for an ideation and design tool that offers an integrated approach to facilitate the process of creating data analogies. 

Recently, the emergence of Large Language Models (LLMs), such as GPT-3~\cite{brown2020language}, has expanded beyond the confines of domain-specific datasets and could efficiently provide more diverse inspirations. Meanwhile, Text-to-Image Generative Models, such as Stable Diffusion~\cite{rombach2022high}, offer an additional efficiency dimension in producing visual representations for data analogies. These advanced Generative AIs possess the capability to generate design ideas and produce illustrative materials based on the generated text descriptions, thus streamlining the data analogy creation process.

However, using generative AI techniques to assist in the data analogy creation process presents several challenges. 
First, there is a lack of systematic design space for data analogies. Such a design space could serve as a guiding framework for large language models in their design process, ensuring that they could generate contextually relevant and insightful analogies. Second, crafting a data analogy solution involves multiple steps. It requires a nuanced understanding of the underlying meaning of numbers and objects and several design considerations, such as the style of illustrations, color palettes, and their alignment with the thematic context. To streamline the intricate process, it is imperative to develop a systematic pipeline to generate analogy solutions step-by-step. 
This pipeline should encompass the important stages of data analogy creation, from concept development to visual representation designs, ensuring a cohesive and efficient workflow.
Last, to facilitate the creation process and provide a user-friendly experience, we need to carefully design tailored user interactions while integrating different Generative AI techniques into such a creativity support tool for data analogies.

\begin{figure}
    \vspace{-2mm}
    \centering
    \includegraphics[width=0.9\textwidth]{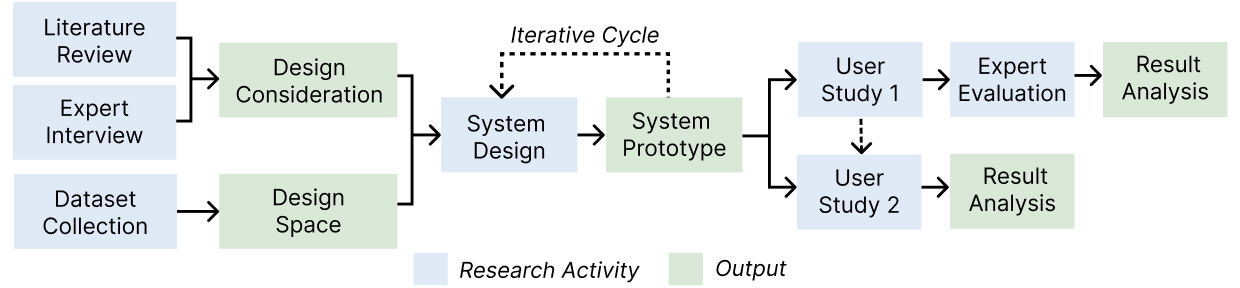}
    \vspace{-2mm}
    \caption{\rv{The research process of the paper, where the blue boxes are different research activities while the green boxes are the output of such activities.}}
    \vspace{-2mm}
    \label{fig:researchStep}
\end{figure}

In addressing these issues, this paper embarks on a mission to leverage data analogy to bridge the gap between abstract data and familiar measurements, thereby \rv{facilitating} data comprehension and enhancing data communication.
\rv{We elaborate on our iterative research process following the framework illustrated in Fig.~\ref{fig:researchStep}.} 
We first conduct semi-structured interviews with design experts to identify the problems in design practices and derive design considerations from the feedback. 
Then, we collect exemplars from various online sources to characterize the design space when creating illustrations of data analogies. 
Next, we propose a data analogy generation pipeline and develop a prototype system that automatically suggests data analogy choices with their corresponding design solutions. The system also supports generalizing detailed images powered by generative AI to illustrate the analogies. 
\rv{Finally, we conduct two user studies. The first user study is a within-subject lab study with 16 amateur designers to evaluate the utility of our prototype system in facilitating the creation of data analogies. The second user study is a crowdsourcing task with 80 participants to examine the effectiveness of the created data analogies using AnalogyMate in improving data comprehension. }
\rv{ 
The major contributions of the paper are as follows:
\begin{itemize}
    \item A design space for data analogies based on the collected dataset of 138 classified data analogy cases.  
    \item A creativity support tool called AnalogyMate that facilitates the creation of data analogies using generative AI, and two user studies that demonstrate its effectiveness in enhancing data comprehension and communication.  
\end{itemize}
}

\section{Related Work}
\subsection{Data Analogy}
\rv{The importance of data is increasingly evident in both scientific and societal contexts~\cite{kjelvik2019getting}, underscoring the crucial need for a profound understanding, precise analysis, accurate interpretation, and effective communication of data results~\cite{muniz2020deficits}.}
Unfamiliar measurements or numbers are frequently encountered in textual content, often posing challenges for readers due to their lack of immediate meaning without the appropriate contextual cues~\cite{barrio2016improving}. 
\szd{A strategic way to improve how well people understand and communicate data is to use data analogy, which connects abstract data to objects or measurements that people are familiar with. This technique addresses the challenge of making data more accessible, as}
Riederer et al.~\cite{riederer2018put} provided empirical evidence showcasing the pivotal role of analogies in facilitating numerical comprehension and recall. 

\szd{From a theoretical standpoint,} Chevalier et al. \cite{chevalier2013using} proposed a framework to facilitate the comprehension of quantitative information through visual representations. Barrio et al.~\cite{barrio2016improving} extended this by introducing a set of perspective templates, specifically designed to convey unfamiliar measurements to a broad audience across formats. \rv{Riederer et al.~\cite{riederer2018put} introduced the concept of ``re-expression'' using measurements of concrete objects for unfamiliar measurements. }
\rv{These approaches employ the re-expression method, substituting familiar measurements for less accessible ones, whereas the perspective template approach distills data into simpler forms using percentages, ratios, and multiples. The data analogy technique synthesizes both methods, incorporating elements of both to transform complex information into more comprehensible terms.}
Despite these valuable contributions, prior research had yet to categorize data analogies systematically. To fill this gap, this paper introduces a design space of data analogies by systematically classifying 138 cases collected from various prominent websites. 

Meanwhile, some researchers explored automatic methods for generating data analogies. Hullman et al. \cite{hullman2018improving} pioneered the creation of a comprehensive database encompassing familiar objects and \szd{their corresponding} measurements. 
\szd{They also created an innovative tool capable of translating less known measurements into terms of familiar objects' measurements. For example,}
Kim et al.~\cite{kim2016generating} introduced a tool for generating personalized spatial analogies to help readers understand distances and areas in text articles. However, traditional automatic analogy generation tools relied heavily on curated data sets \rv{while we utilized the capabilities of GPT-3.5 to assist users in analogy designs. Traditional methods are typically confined to analogies within the same measurement, whereas our approach enables measurement transformation. Additionally, we employ stable diffusion to generate illustration materials based on design schemes.}


\subsection{Creativity Supporting Tool}
Creativity was once considered a unique skill that makes humans distinctive~\cite{sawyer2011explaining}. 
However, with significant advancements in LLMs and natural language processing (NLP) techniques,  \rv{the utilization of artificial intelligence to construct Creativity Support Tools (CSTs, e.g., \cite{wan2023gancollage, wang2010idea}), has emerged as a trend. These tools facilitate a spectrum of tasks performed by designers~\cite{kim2022mixplorer}, physicians~\cite{zhou2023beyond}, and researchers ~\cite{portenoy2022bursting}.}

\rv{In the realm of AI-supported CSTs, two major groups could be delineated based on the specific aspect of the process they support.
One group of CSTs focuses on background research and ideation evaluation~\cite {frich2019mapping}. 
For instance, Zhang et al.~\cite{zhang2022storydrawer} have developed \textit{Story Drawer}, a cooperative drawing platform between children and AI that aims to promote children’s imaginative visual storytelling. 
Wang et al.~\cite{wang2010idea} designed \textit{Idea Expander}, which facilitates group brainstorming by providing a dynamic visual communication channel and pictorial stimuli.
Moreover, Karolus et al.\cite{karolus2023your} generated meaningful task-related proficiency feedback to improve user awareness of the issues in their writing samples.}

\rv{The other group of CSTs aims to assist the designing process by directly generating new ideas or giving hints on the target topics~\cite{frich2019mapping}. 
Wan et al.~\cite{wan2023gancollage} proposed a StyleGAN-driven digital mood board, \textit{GAN-Collage}, integrating AI-generated visual ideas into the ideation phase to support creativity. 
Mozafari et al.\cite{mozaffari2022ganspiration} introduced an approach for both targeted and serendipitous UI design inspiration on image-based inputs.
Moreover, some researchers leverage metaphor and analogy to find connections between ideas that share similar characteristics. 
For example, Sun et al. proposed using metaphor's target-source concepts to facilitate visual design ideation~\cite{10.1145/3411764.3445325} and gesture design~\cite{10.1145/3511892} imaginatively. 
To support science writers in explaining unfamiliar concepts, Kim et al.~\cite{kim2023metaphorian} launched \textit{Metaphorian}, which facilitates the search, extension, and iterative revision of extended metaphors.}

The current trend of CSTs could be summarized as low maturity, all-process support, multi-device compatibility, low complexity, low availability, and low expertise~\cite{frich2019mapping}. 
\rv{Our work focuses on facilitating the entire creativity process, by highlighting generating, evaluating, and utilizing ideation in analogy designs and their downstream tasks.}
Building on the concept of inspiring creativity in CST design~\cite{reed2023negotiating,krauss2022elements}, our tool supports the designer and authors in performing end-to-end analogy generation tasks through analysis, revision, and utilization of inspirations. 

\subsection{LLM-Supported Design}
The emergence of Large Language Models(LLM) such as GPT-3~\cite{brown2020language} and Large-scale Text-to-image Generation Models(LTGM) like Stable Diffusion~\cite{rombach2022high} has spurred a wave of research in the Human-Computer Interaction (HCI) domain. Researchers have been investigating the integration of advanced artificial intelligence technologies to support various tasks. These tasks span a wide spectrum of HCI applications, encompassing question-answering systems~\cite{ma2023demonstration}, story generation~\cite{dang2023choice,chung2022talebrush}, automatic generation of visualizations~\cite{dibia2023lida} and personalized news recommendation~\cite{liu2023first}. For example, CodeToon~\cite{suh2022leveraging} is a visual programming environment that fosters collaborative comic creation through generative conversational AI. Opal~\cite{liu2022opal} leads users in a systematic exploration of visual concepts and offers a pathway for generating illustrations based on article tone, keywords, and relevant artistic styles using semantic search and methods of prompt engineering GPT-3. 


In the context of pre-trained generative models, prompt engineering methods offer an alternative approach to model fine-tuning~\cite{wang2023reprompt}. Recent work in prompt engineering has proposed several methods to improve prompts~\cite{haviv2021bertese, jiang2020can}. Specifically, the development of Low-code LLM~\cite{cai2023low} and PromptCrafter~\cite{baek2023promptcrafter} have been dedicated to capturing user intent through well-crafted text prompts and optimizing the wording of prompts. Dang et al.~\cite{dang2023choice} integrated a user interface (UI) for phrase suggestions and a UI for zero-shot prompt inputs into an LLM. Our approach integrates the chain-of-thought technique~\cite{wei2022chain} and leverages few-shot examples~\cite{brown2020language} to enhance the prompt optimization, ultimately aiming for improved and more effective outcomes.
In our study, we employ GPT-3.5 as a knowledge repository to assist users in interactively designing data analogies and illustrations. Additionally, we offer prompting solutions that establish connections between Large Language Models and Large-scale Text-to-image Generation Models.

\section{Preliminary Study}
To better understand the workflow and design choices in creating data analogies, we conducted in-depth interviews with four experienced domain experts and summarized the design considerations from expert feedback.

\subsection{Expert Interview} 
We conducted semi-structured interviews with four design experts (E1-E4, two females) following established research guidelines~\cite{lazar2017research}. Each interview lasted approximately one hour and was conducted online via meeting software that enables screen sharing. All the interviewees have over 7 years of experience in data-related design and have personally created or frequently encountered data analogies. We first introduced our research topic and the core concepts (i.e., data analogy) to the experts. Next, we asked them a set of prepared questions under two topics: the usage scenarios of data analogies and the general workflow for creating data analogies. Example questions included queries such as, ``\textbf{what are the common use cases, and what motivates their adoption?}'', ``\textbf{could you outline the typical creation process for data analogies?}'', and ``\textbf{what factors and design requirements should be considered during the creation process?}''. Throughout the interview process, we asked follow-up questions if we noticed that an interviewee's answer was unclear or if we wanted to dig deeper into the details of their daily conduct. 

\rv{
We collected a substantial amount of qualitative data from the interviews. We transcribed the audio recordings and then coded the texts to achieve our interview objectives (design workflow and design consideration) through thematic analysis \cite{braun2006using}. Two authors were responsible for the coding process. Initially, we independently coded the transcriptions, marking sentences relevant to our research questions. Subsequently, for each research question, we read through all marked sentences, generated codes from them, and grouped similar codes. After independent coding, we met to compare our codes and discussed any discrepancies until we reached an agreement. We then coded all transcriptions using the latest codes. Our findings are summarized into two parts: design workflow of data analogy creation, and design considerations for building a creativity support system.}

\rv{
\subsection{Design Workflow}
During the interview process, we identified a common need to provide a systematic design pipeline from all the experts. The pipeline of data analogy creation can be typically divided into two sequential stages: analogy design and illustration design. We summarize the related expert feedback as follows.

\underline{\textit{Provide a systematic pipeline.}}
All four experts mentioned that establishing a well-organized analogy design process is necessary and beneficial. E1 expressed the desire for users to engage in interactive decision-making at crucial moments. E2 also shared his ideal workflow, stating that \textit{``I hope the system can provide automatic recommendations for different analogy solutions and design schemes.''} 

\underline{\textit{Facilitate analogy design.}}
All four experts agree that analogy design is a time-consuming process that requires domain knowledge and iterative refinement. E1, E3, and E4 emphasize that analogy design demands sensitivity to data, an understanding of conceptual nuances within the data, and relevant domain expertise. E1 suggests, \textit{``Data analogy design typically involves systematic brainstorming to find the most fitting metaphorical expressions.''} 

\underline{\textit{Assist illustration design.}}
Two experts (E1 and E2) believe that the visual design of data analogy illustrations is crucial. E2 shared his personal experience of spending a significant amount of time searching for suitable image materials, stating that \textit{``the main challenge lies in the conceptualization of the design process.''} E1 added, \textit{``it is important to align the color palette of the illustrations with the style of the article.''}
}


\rv{
\subsection{Design Cosideration}
Combined with expert interviews and our previous literature review, we have derived the following design considerations for the proposed system, specifically aimed at addressing their key concerns.}

\rv{
\textbf{C1. Understand the implications of the data content.}
In designing data analogies, it is crucial to help the audience quickly grasp the underlying implications of the data content descriptions. As mentioned by E3, \textit{``the goal of designing data analogies is not merely to present the data but to encourage readers to engage in deeper contemplation.''} 
}

\rv{
\textbf{C2. Enhancing data understanding through analogy designs.} Based on previous literature and expert feedback, we have summarized four key factors influencing data comprehension: object familiarity, concreteness, similarity~\cite{hullman2018improving, kim2016generating}, and the perceptibility of numerical values~\cite{bautista2011make}. For the first three factors, we aim to leverage user-friendly interactions, allowing users to customize their preferences according to their needs. 
Concerning perceptibility, we also considered this when designing prompts. The specific implementation of the four key factors can be found in Section 5.3.
}

\rv{
\textbf{C3. Ensure the consistency of illustration styles with the topic domain.}
All the experts concurred that the design of illustrations should align with the thematic essence conveyed by data analogies. For instance, if a data analogy addresses a serious subject, the design of the illustrations should exude a sense of seriousness. Maintaining a consistent style across multiple illustrations is imperative. Illustration design involves considering numerous factors, encompassing emotions, stylistic elements, and color palettes. Our system adheres to various design principles to ensure the coherence of visual elements with the narrative topic throughout the design process.}

\rv{
\textbf{C4. Enable user control for customization and personalization.}
All experts expressed their preferences for controlling the generated results at various stages. First, they wish to have control over different analogy strategies to explore different analogy ideas. Second, when designing for different scenarios, they hope to adjust parameters to meet specific requirements. After the system generates automatic analogy descriptions, they are more inclined to review and modify the generated analogies for fine-tuning. Finally, when creating final illustrations using generative models, users would like to adjust keywords or prompts for better control over the generated illustrative materials.}

\section{Design Space}

\rv{Previous studies lacked a systematic design space of data analogies. To provide guidance for our system, we collected a dataset of 138 cases from online sources. Based on a systematic classification of these cases, we proposed the design space for creating data analogies.} 


\subsection{Dataset Construction}
In our data collection and labeling process, two researchers with different backgrounds (one majoring in HCI and one with a data science background) were involved. To ensure a standardized procedure, they carried out the data collection, labeling, and screening process separately. 
We initiated our study by conducting a comprehensive search for relevant datasets that had been previously released in academic research and other sources, including infographics~\cite{borkin2013makes} and data videos~\cite{yang2023understanding}. 
These databases were sourced from reputable sources (e.g., \textit{The Washington Post}, \textit{Reuters}) and were of high quality (e.g., winners of \textit{Kantar Information is Beautiful Awards}). 
To compile a diverse set of analogy cases, we then collected our analogy exemplars by querying and browsing multifacet data-related online sources. 
Our source website covers various types of media, including news articles (30\%), social media (20\%), infographic websites (30\%), and institute websites (20\%). After identifying the analogy cases from the websites, we deconstructed them into analogy pairs.
After the dataset was ready, the two researchers conducted the data labeling process independently.
They labeled each case with predefined data types and representation styles.
Inter-rater reliability was quantified by Cohen's kappa coefficient: 0.95 (analogy strategy), 0.97 (measurement transformation), 0.82 (data binding types), 0.98 (presentation form) and 0.98 (layout).
For the inconclusive cases, two researchers proceeded with the screening process by first extracting the controversial ones. Subsequently, they engaged in a discussion to assess whether these cases met the definition of data analogy and eliminated any unqualified items.
The resulting data analogy cases were compiled into an Excel file, where any duplicates or unachievable cases (i.e., protected by property rights or refused to be reprinted) were removed. 
Finally, after all the data were ready, two researchers resolved the differences of opinion arising from the collection and labeling processes through further discussions.





The final dataset consists of 138 analogy cases, totaling 3,178 analogy pairs, which were extracted from 66 different online pages.
The topic of these analogy cases covers fields of health (15.4\%), entertainment (23.1\%), ecology (10.1\%), popular science (23.3\%), and society (18.5\%). \rv{The dataset is available at \url{https://analogymate.github.io/designspace/}.}

\begin{figure}
    \vspace{-1mm}
    \centering
    \includegraphics[width=1\textwidth]{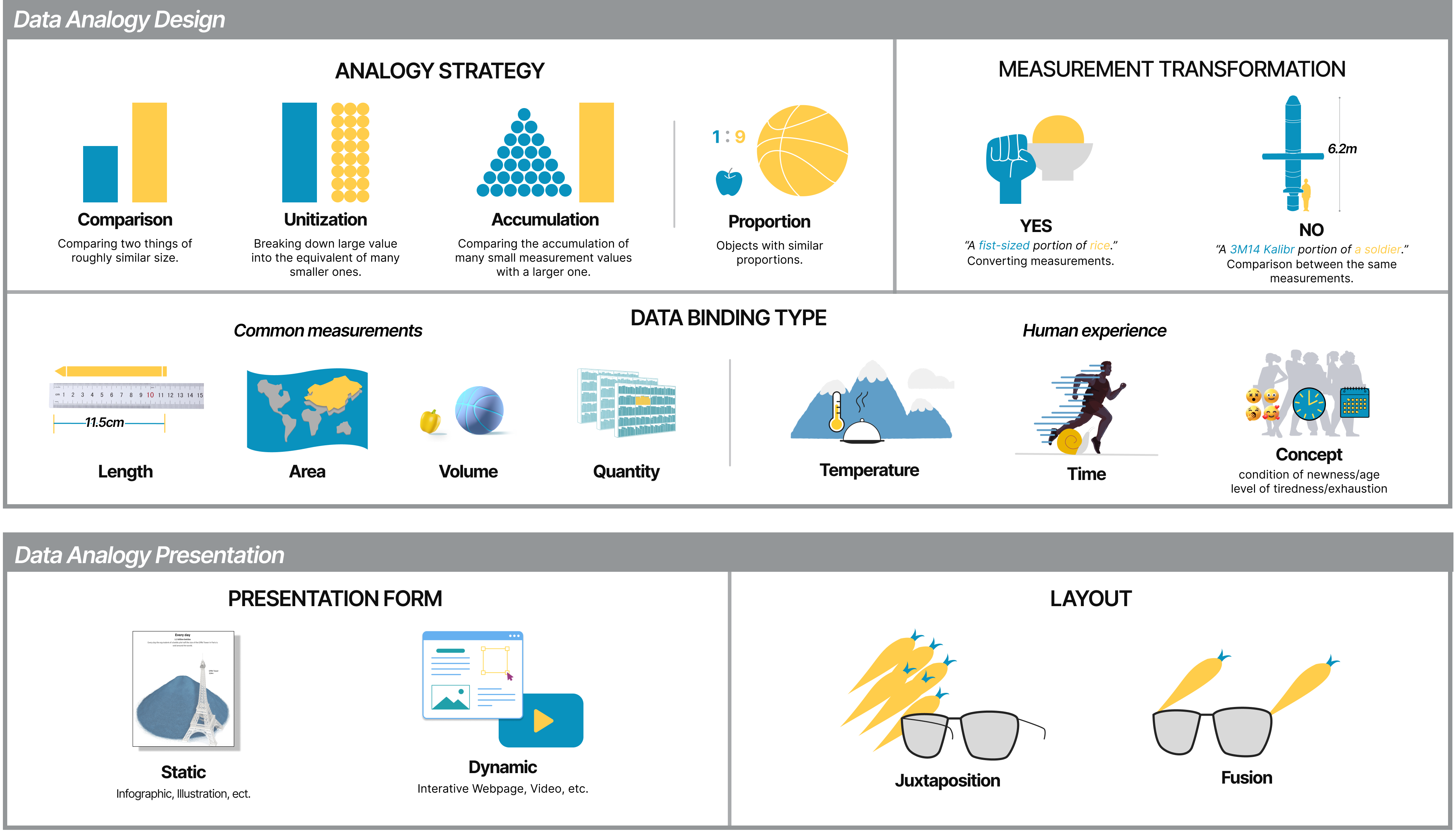}
    \vspace{-6mm}
    \caption{Data analogy design space, which consists of two perspectives: data analogy design (analogy strategy, measurement transformation, and data binding type) and data analogy presentation (presentation form and layout).}
    \vspace{-4mm}
    \label{fig:DesignSpace}
\end{figure}

\subsection{Data Analogy Classification}

Our deconstructive analysis of the collected data serves two primary purposes: to classify data analogy cases and to enhance our understanding of how concepts are conveyed through various relationships between objects and entities. We then construct our design space from two perspectives: data analogy design and data analogy presentation (as shown in Fig.~\ref{fig:DesignSpace}). The former relates to the data analogy ideation process, including analogy strategies, measurement transformation, and data binding types. The latter corresponds to the data analogy visual presentation process, including presentation form and layout.

\subsubsection{Data Analogy Design}



To understand how data analogies are created, we surveyed previous literature and analyzed the data analogy dataset we collected. We then characterized three key elements of data analogy design: analogy strategy, measurement transformation, and data binding type. 

\begin{figure}
    \centering
    \includegraphics[width=0.9\textwidth]{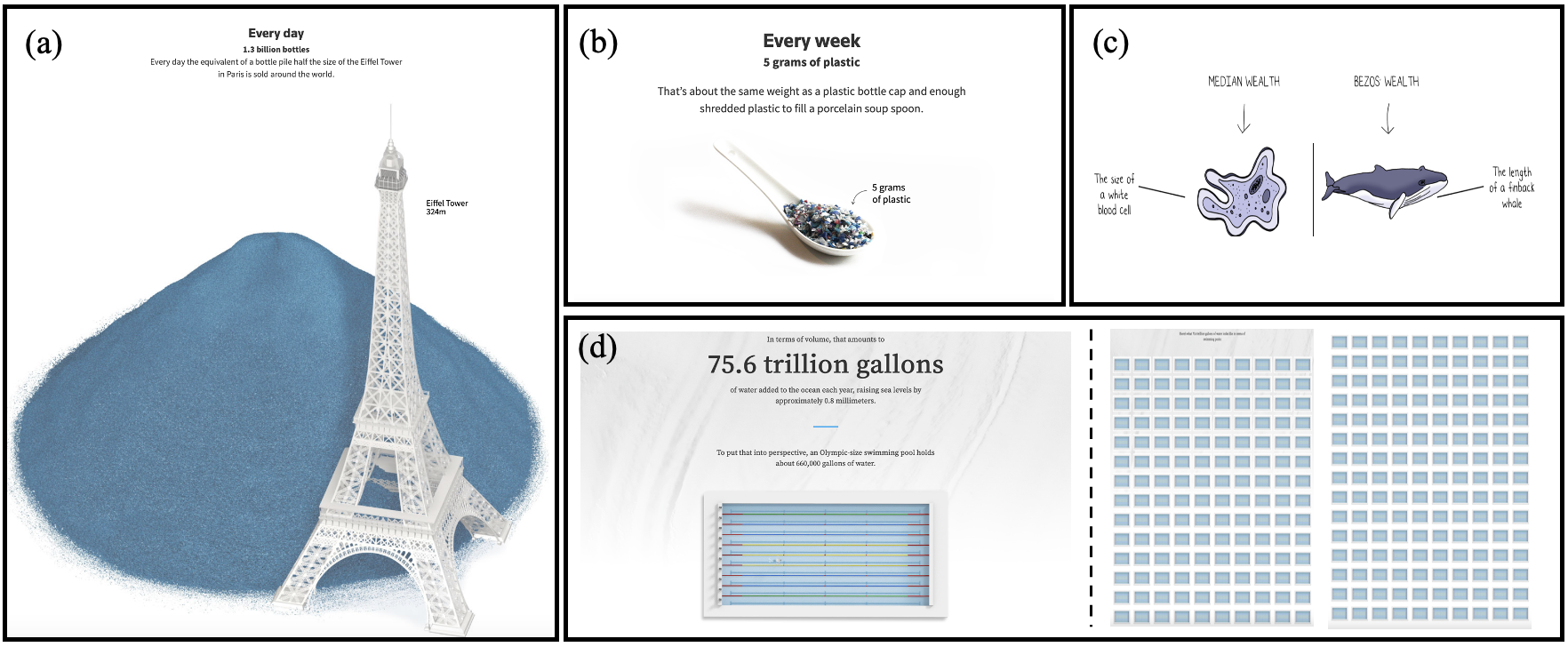}
    \vspace{-2mm}
    \caption{Four data analogy examples from our collected dataset. (a) 1.3 billion bottles are sold daily compared with the \textit{Eiffel Tower}. (b) Five grams of plastic are consumed weekly compared to a porcelain soup spoon. (c) The wealth difference between \textit{Jeff Bezos} and an American mid-class family compared with the size difference between a white blood cell and a finback whale. (d) Each year, 75.6 trillion gallons of water are added to the ocean, which equals 114.4 million Olympic-size swimming pools.}
    \vspace{-4mm}
    \label{fig:DataAnalogyExample}
\end{figure}

In terms of \textbf{analogy strategy}, Chevalier et al.~\cite{chevalier2013using} demonstrated 3 object-to-object relations with regard to their measurable properties on a scale: comparisons, containment, and unitization. \textit{Comparison} (Fig.~\ref{fig:DataAnalogyExample}(a)) focuses on the elementary-level shared properties between objects. This aligns with one-to-one type analogies, where objects are directly compared with another analogical object. 
\textit{Containment} refers to objects placed within a container. 
\textit{Unitization} (Fig.~\ref{fig:DataAnalogyExample}(b)) involves redefining an object as a new unit of measurement.  
Similar to the \textit{untization} strategy, we proposed another strategy, \textit{accumulation}, which is also commonly applied in analogy designs. When the scale of the original object is too small to grasp, \textit{accumulation strategy} (Fig.~\ref{fig:DataAnalogyExample}(c)) is employed to present a bunch of the original objects to form another analogical object, which is the opposite operation as in \textit{unitization}. 
\rv{Hullman et al.~\cite{hullman2018improving} defined proportional analogy as expressing a pair of measurements using two familiar objects that have measurements with the same ratio, and we also adopted \textit{proportion}(Fig.~\ref{fig:DataAnalogyExample}(c)) strategy.}
In conclusion, we have expanded the classification of data analogy strategies, which include comparison, unitization, accumulation, and proportion. By analyzing our dataset, we found that the most commonly used strategy is comparison (73.4\%), followed by accumulation (13.7\%), proportion (7.2\%), and unitization (5.8\%). 

In terms of \textbf{measurement transformation}, we found that 
for some measurements like temperature, time, and wealth, it is sometimes difficult to design proper analogies within the same data measurements. For example, Fig.~\ref{fig:DataAnalogyExample}(c) shows an analogy example using measurement transformation, in which the wealth gap between Jeff Bezos and a typical middle-class family is measured as the size of a white blood cell and a finback whale. The transformation process, or the switch of measurement, conveys a more intuitive message about the given objects. 
In our dataset, a significant 67.6\% of analogical cases involve measurement transformation. 

In terms of \textbf{data binding type}, we have identified seven common data binding types used in data analogies. Four of the seven data binding types have been included in previous research~\cite{shi2022supporting}, namely length, area, volume, and quantity. These four types are also common measurements that are easily depicted in static illustrations. The other three types, temperature, time, and abstract concepts, are derived from our collected dataset. All three additional data binding types are highly related to human experience. 
By analyzing our dataset, we found that length(50, 35.9\%) stands out as the dominant data binding type, with volume (34, 24.5\%), area (28, 20.1\%), and quantity (14, 10.1\%) trailing behind. Abstract concepts (10, 7.2\%) also play a vital role in pictorial visualizations while time (2, 1.4\%) and temperature (1, <1\%) were rarely utilized in pictorial visualizations. Such design decisions may be due to the relatively low perception of these human experiences.



\subsubsection{Data Analogy Presentation}

To present an analogy, there are various presentation forms, including illustrations, infographics, video, interface website, etc. By analyzing our dataset, we have classified those presentation forms into static and dynamic forms (as shown in Fig.~\ref{fig:DesignSpace}). We observed that the majority of data analogy cases (89, 64.5\%) in our dataset use static forms, while the other cases (49, 35.5\%) apply dynamic forms. 

The layout of objects is another crucial consideration in creating visual representations. Phillips and McQuarrie~\cite{phillips2004beyond} proposed a typology to distinguish different metaphor structures based on the complexity and ambiguity of the visual structure. We adopted two classifications, juxtaposition, and fusion, that are applicable in data analogy scenarios. As shown in Fig.~\ref{fig:DesignSpace}, the juxtaposition layout refers to a graphic in which objects and analogy entities are separated and distributed in parallel, while the fusion layout refers to a graphic in which objects and analogy entities blend harmoniously into one. 
\rv{In our dataset, apart from a few cases containing only text (14, 10.2\%), the juxtaposition layout (82, 59.4\%) has a numerical advantage over the fusion layout (42, 30.4\%).}
\rv{The fusion layout, blending original and analogical objects harmoniously into one design, is more challenging and time-consuming than the juxtaposition layout. This challenge arises from the need to achieve a seamless integration and coherent visual representation.}

\section{System}
\begin{figure}
    \centering
    \includegraphics[width=1\textwidth]{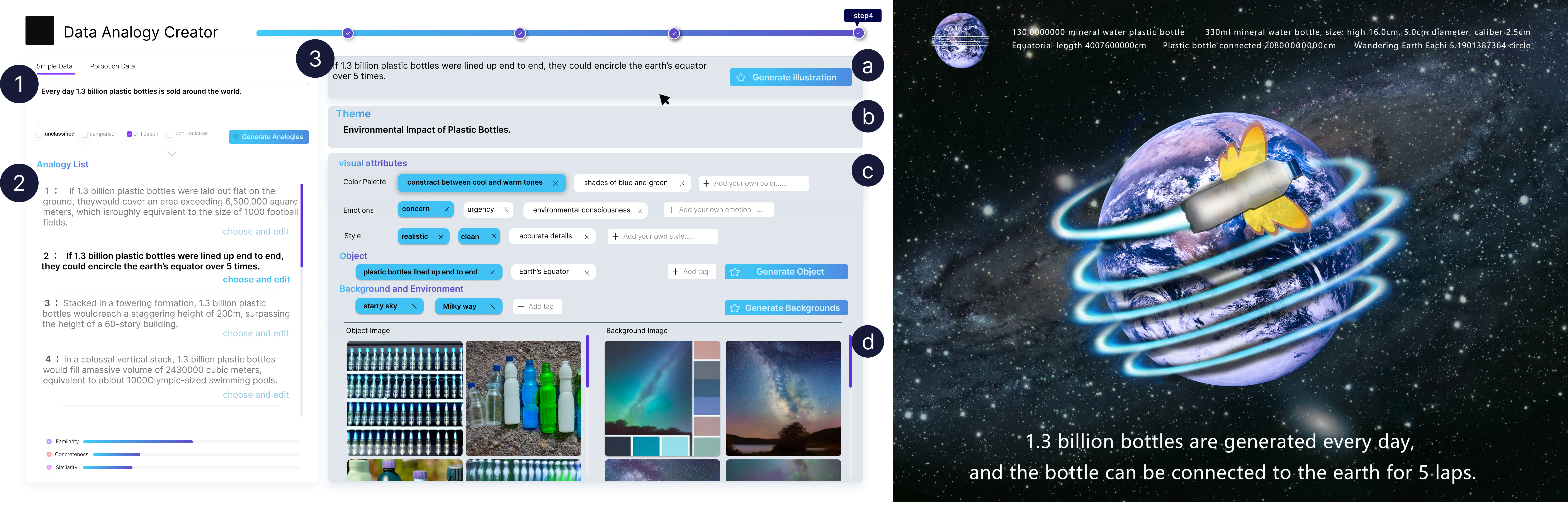}
    \caption{System interface (left) and illustration created by users using generated analogy and materials (right). The system in this screenshot has suggested an analogy list for the input ``Every day 1.3 billion plastic bottles are sold around the world'' and generated illustration design solutions for the chosen analogy.}
    \vspace{-4mm}
    \label{fig:interface}
\end{figure}

To support creative design with data analogy, we developed a prototype system called \textit{\systemname} through an iterative process. The system (Fig.~\ref{fig:interface}) employs an automated two-stage pipeline (Fig.~\ref{fig:pipeline}) to suggest analogy design and illustration design. 
In the analogy design process, the underlying meaning of the sentence is interpreted to suggest analogy objects better match the theme (\textbf{C1}), and four factors are taken into consideration to enhance data comprehension through generated design solutions (\textbf{C2}). In the illustration design process, the system suggests different aspects of the illustrations according to the theme (\textbf{C3}). Also, users can make crucial decisions interactively at critical points throughout the process (\textbf{C4}). 

\subsection{Interface Design and User Interaction}
{\systemname} consists of three main views: the Input View, the Generator View, and the Refinement View.
\textbf{Input View (V1)} allows users to input original data description sentences containing numerical data. Users can select different design strategies according to their requirements.
\textbf{Generator View (V2)} automatically suggests data analogies, their corresponding design solutions, and illustration materials.
\textbf{Refinement View (V3)} is where users can sort the analogy list to meet their specific needs and make adjustments to the generated results.

Here is how the system works. 
Users begin by entering the Input View (depicted in Fig.~\ref{fig:interface}-1), where they input description sentences containing numerical data. Users can choose between simple data and proportion data as their input formats. For simple data, users have the option to choose from three analogy strategies: comparison, unitization, and accumulation. If users are unsure about which analogy strategy to use, they can select the ``unclassified'' button, allowing the system to generate various types of analogies using all the above strategies. 
Next, users click the ``Generate Analogies'' button, and the system returns a sorted list of generated analogies, as shown in Fig.~\ref{fig:interface}-2. Users can update the sorted list by adjusting the weights of three factors using the slider bars at the bottom, aligning the results with their preferences. 
Subsequently, users can select one of the analogies and click on the ``choose and edit'' button. The selected analogy result then appears in the text area shown in Fig.~\ref{fig:interface}-3(a). Here, users can make modifications to the analogy description and then click the button to generate an illustrated design scheme, which includes the interpretation of the data analogy theme (Fig.~\ref{fig:interface}-3(b)) and the recommended illustration keywords (Fig.~\ref{fig:interface}-3(c)). Among those keywords, ``visual attributes'' include keywords related to the illustration's emotion, style, and color palette, while ``objects'' and ``background'' respectively represent the recommended illustration objects and background. Users can add or modify keywords and select multiple keywords to generate illustration materials (Fig.~\ref{fig:interface}-3(d)).

\subsection{System Pipeline}

We introduce an automated two-stage pipeline (Fig.~\ref{fig:pipeline}) to construct the {{\systemname} system, tailored for creating data analogies with analogy designs and corresponding illustration materials. 
In the first stage, the system proceeds through three steps to generate data analogies. In the second stage, it generates illustration design schemes along with the corresponding illustration materials.

\subsubsection{Stage 1: Analogy Design}
This first stage consists of three major steps: generate analogy objects, modify generated analogy objects, and calculate and polish the descriptive sentence.

\underline{\textit{Generate Analogy Objects.}} When a user inputs a sentence containing numerical data into {\systemname}, the system generates familiar objects with their measurements according to the chosen analogy strategies. This process is accomplished using GPT-3.5, which involves three specific steps: prompt selection, guided design, and theme interpretation.
To design prompts, we start by selecting representative data analogy examples as few-shot examples. After that, based on the user's chosen analogy strategies, we provide corresponding guidelines to guide the design process. Finally, to interpret the underlying theme of the input, we utilize both few-shot examples and the chain-of-thought method. 

\underline{\textit{Modify Generated Analogy Objects.}} To enhance the data accuracy and clarity of generated analogy objects, we employ GPT-3.5 to perform a two-step correction. In the first step, we verify and adjust the numerical values of the analogy objects. This step enables us to correct obvious errors, such as changing ``depth of swimming pool: 0.2 meters'' to ``depth of swimming pool: 2 meters''. In the second step, we address issues where data descriptions lack clarity. For example, we may modify ``the height of a house: 7 meters'' to ``the height of a two-story house: 7 meters.'' This further refines and optimizes the generated analogy objects by providing more specific measurement values.

\begin{figure}
    \vspace{-2mm}
    \centering
    \includegraphics[width=1\textwidth]{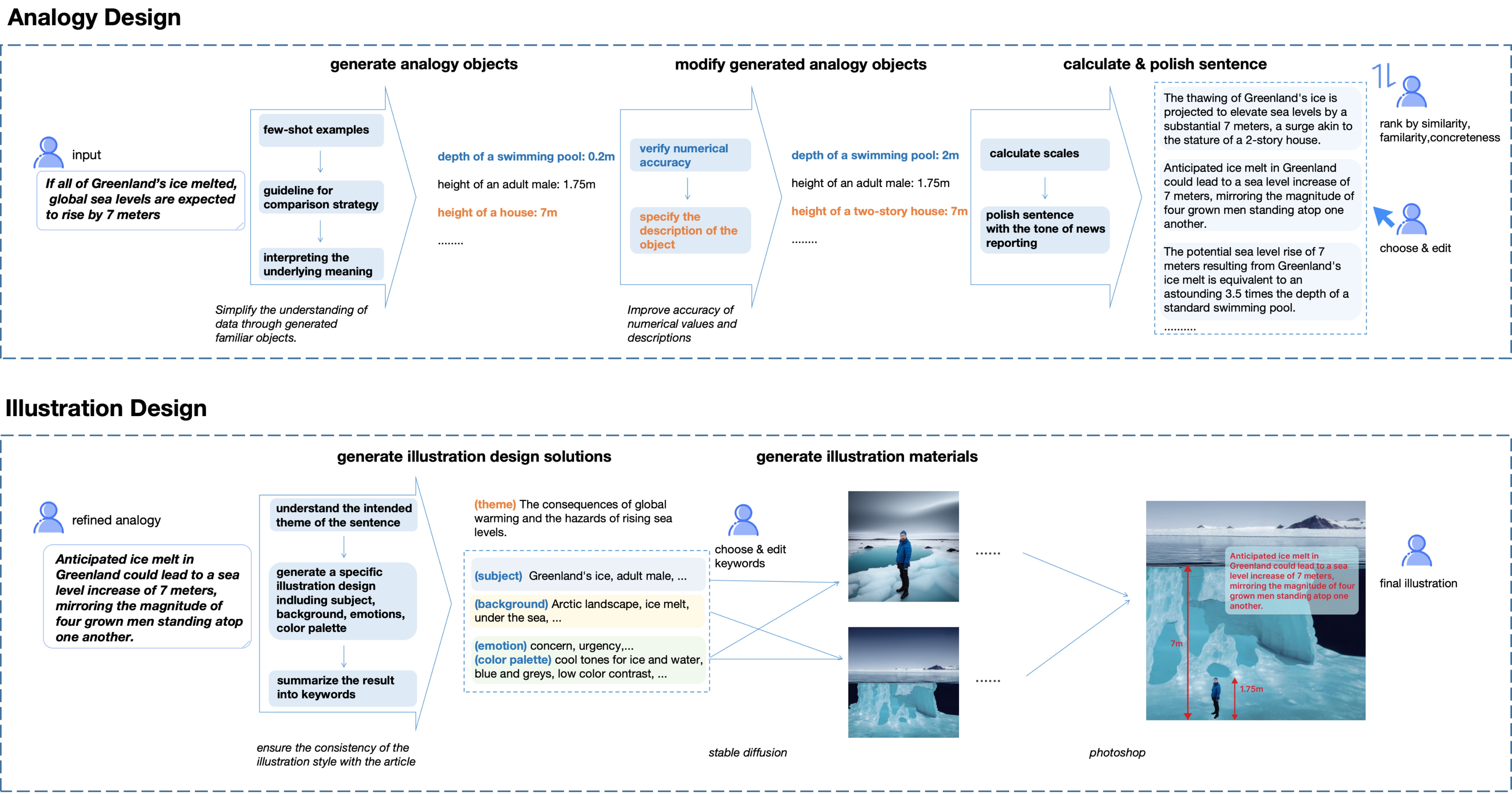}
    \vspace{-6mm}
    \caption{Pipeline of {\systemname}. In the analogy design process, the system first generates analogy objects, then modifies the results, and finally calculates and polishes the analogy sentence. In the illustration design process, the system suggests illustration design solutions and then generates illustration materials according to the keywords selected. Users can make crucial decisions interactively at critical points throughout the process.} 
    \vspace{-2mm}
    \label{fig:pipeline}
\end{figure}

In our experiments, we observed that using GPT-3.5 to calculate generated analogies within the same conversation can lead to calculation errors and inconsistencies between the calculation results and the answers. To address these issues, we opt to perform calculations on our backend server. Then, we construct sentences using predefined text templates. Following this, we utilize GPT-3.5 for sentence refinement using news report tone.
\subsubsection{Stage 2: Illustration Design}
The second stage involves two major steps: generate illustration design solutions and generate illustration materials.

\underline{\textit{Generate Illustration Design Solutions.}} This step also employs the chain-of-thought method (Fig.~\ref{fig:pipeline}). Initially, GPT-3.5 needs to understand the intended meaning of the analogy sentence to ensure that the generated illustration schemes align with the theme and emotions users want to convey. Second, our system suggests different aspects of the illustrations, including visual attributes, objects, and backgrounds. Visual attributes encompass factors such as emotion, color palette, and style. Previous research~\cite{wei2006image} has shown that emotions in images can be conveyed through color factors such as color temperature, brightness, and color contrast. Therefore, when generating color palettes, our system suggests these four factors based on the theme. 
Finally, each aspect of the illustration design results is extracted into keywords.

\underline{\textit{Generate Illustration Materials.}} We utilize the Stable Diffusion API to generate illustration materials. To ensure consistency in style among the generated illustration materials, we maintain constant keywords of visual attributes when generating the main object and background images. This approach facilitates the post-processing and integration by designers, ensuring that the final illustrations maintain a cohesive and harmonious style.

\subsection{Implementation Detail}

This system was implemented using Python's Flask for the backend, while the frontend was developed with React, CSS, JavaScript, and Axios. Additionally, utilized the Stable Diffusion and GPT-3.5 APIs to perform the analogy design and illustration design tasks. 
\rv{To enhance data comprehension, we have incorporated the factors outlined in \textbf{C2} into our system: object familiarity, concreteness, similarity, and numerical value perceptibility. For the initial three factors, our aim is to enable user-adjustable weights, offering customization according to individual preferences. Regarding perceptibility, we consider it when designing prompts.}

\subsubsection{Object-entity Similarity}
The similarity between objects and analogy entities strongly influences the freshness and vitality conveyed by analogy cases. 
In the field of Natural Language Processing (NLP), quantifying similarity between concepts has been achieved, with notable frameworks such as WordNet~\cite{budanitsky2006evaluating} and ConceptNet~\cite{speer2017conceptnet}. To ensure consistency and avoid the impact of different score ranges, we normalized the indicators.
\begin{align}
   \text{Norm}(I(x,y_i)) &= \frac{\text{I}(x, y_i) - \min_{(x,y_j) \in (x,Y)} \text{I}(x,y_j)}{\max_{(x,y_j) \in (x,Y)} \text{I}(x,y_j) - \min_{(x,y_j) \in (x,Y)} \text{I}(x,y_j)}
\end{align}
Note $x$ refers to the object concept of a given data analogy case. $Y$ represents the set of analogy entity concepts. $I(x,y_i)$ is the measurement to be normalized, such as similarity in WordNet.

Accordingly, we define a similarity term to quantify and adjust the semantic relationship between object and entity, given as
\begin{align}
   S_i &= w_1\text{Norm}(\text{Path}(x, y_i)) + w_2\text{Norm}(\text{Corr}(x, y_i))
\end{align}
where $Path(x,y_i)$ measures the length of the shortest path in the WordNet hyponym/hypernym graph~\cite{perkins2010python,hullman2018improving}. $Corr(x, y_i)$ represents relatedness defined in ConceptNet, which exhibits a high correlation with the human gold-standard ranking, as measured by its Spearman correlation~\cite{speer2017conceptnet}.

\subsubsection{Familiarity}
Familiarity is a vital factor to measure in analogy design~\cite{hullman2018improving}. A familiar analogy entity provides the audience with abundant information to relate immediately to their own experiences.
While WordNet does not directly measure familiarity, many researchers use the frequency of each synset's occurrence in a semantic concordance as a familiarity score~\cite{kwankajornkiet2016automatic,cohen2021tackling}. 
 Therefore, we include the word frequency database as a component of our familiarity measurement. 
 \rv{We use SUBTLEX-UK~\cite{van2014subtlex} for word frequency statistics, which is a widely accepted database in the fields of linguistics and natural language processing research.}
 We define our familiarity measurement as follows:
\begin{align}
   F_i &= w_3\text{Norm}(\text{Syn}(y_i)) + w_4\text{Norm}(\text{Freq}(y_i))
\end{align}
where $Syn(y_i)$ represents the number of synsets in WordNet related to concept $y_i$, and $Freq(e_i)$ denotes the word frequency based on the SUBTLEX-UK word database. $w_3$ and $w_4$ are user-adjustable parameters that determine the weight of these factors in the familiarity measurement.

\subsubsection{Concreteness}
Concreteness evaluates the perceived level of an entity or concept. Hullman et al.~\cite{hullman2018improving} refers that concrete metaphors should be objects with physical forms. We quantified the concreteness scores using WordNet~\cite{budanitsky2006evaluating} and collected scores from human-evaluation-based database~\cite{brysbaert2014concreteness}. 
 Feng et al.~\cite{feng2011simulating} and Bolognes et al.~\cite{bolognesi2020abstraction} suggested a correlation between hypernyms, hyponyms, and concept concreteness. This inspires us to use hypernyms and hyponyms as a term to measure concreteness. Therefore, we define our Concreteness term as 
\begin{align}
   C_i &= w_5\text{Norm}(\text{hypernyms}(y_i)/\text{hyponyms}(y_i)) + w_6\text{Norm}(\text{Conc}(y_i))
\end{align}
where $Conc(y_i)$ represents the concreteness score in Brysbaert findings~\cite{brysbaert2014concreteness}. $\text{hypernyms}(y_i)$ and $\text{hyponyms}(y_i)$ denote the number of hypernyms and hyponyms linked with word or concept $y_i$. Both scores are weighted by $w_5$ and $w_6$.

\subsubsection{Perceptibility of Numerical Values}
In our preliminary study, E1 emphasized the importance of numbers being easily perceptible to readers when reading articles. Previous systems encouraged multipliers within the range of 1 to 10 for analogy generation~\cite{hullman2018improving}, we adopted this range in our study. Regarding the unitization strategy, E1 stated, ``unitization is often used to convey a large value for impactful delivery, and readers are sometimes forgiving of precision discrepancies.'' Therefore, we limited the multiplier to values exceeding 1000.

\section{Evaluation}

We conducted two user studies to evaluate the validity and effectiveness of our proposed method and prototype system. 
\rv{In the first user study, we examined whether and how {\systemname} contributes to the design of data analogies and the creation of the corresponding illustrations.
The second user study aimed to examine the effectiveness of the data analogy created by {\systemname} in enhancing data comprehension. }

\rv{\subsection{User Study I: Creating Data Analogies with \systemname}}
To gain insights into whether and how {\systemname} contributes to data analogy design and related illustration creation, we conducted a within-subject study, comparing {\systemname} against the conventional web search results. 

\subsubsection{Experiment Design}
%
As prior research has indicated~\cite{petridis2021symbolfinder}, web search engines serve as a powerful tool for generating ideas and creating visual materials. Designers often utilize Internet searches to help find data analogy inspirations and related illustration materials as raw materials. Specifically, we assessed whether users could (1) generate more data analogy designs and (2) improve the efficiency of illustration production with {\systemname}. Additionally, we evaluated our tool's ability to assist users engaged in creative work through a Creativity Support Index (CSI) evaluation questionnaire~\cite{10.1145/2617588}.

\subsubsection{Participants}
We recruited 16 amateur designers through social media and word-of-mouth (P1~P16, 12 females, with an average age of 22.5 years and a standard deviation of 3.37). They come from diverse backgrounds, primarily students, including both undergraduate and graduate students. Additionally, some work in the internet industry, while others are research assistants. The recruitment criteria were that they all had prior experience in design activities, possessed a basic proficiency in design software, and expressed an interest in creating data analogy illustrations. These participants would receive a commemorative gift as a reward for completing the tasks. They were also allowed to receive two additional awards (one creativity award and one design award) after we collected all the user study results, which the domain experts would select.

\subsubsection{Procedure and Tasks}
During the study, participants were tasked with creating data analogies and producing illustrations, with one utilizing the baseline (internet search) and the other employing the {\systemname} system. Initially, participants were randomly assigned two out of four descriptive sentences with numerical content, which were as follows: ``If all of Greenland's ice melts, global sea levels are expected to rise by 7 meters'', ``In order to rescue the children trapped in the cave, firefighters pumped out 1.2 billion liters of floodwater'', ``Every day, 1.3 billion plastic bottles are sold around the world'', and ``The ratio of Bezo's wealth to those of middle-class American families is about 900,000 to 1''. For the assigned sentences, they were required to develop as many data analogy design schemes as possible within a specific time limit. 
Subsequently, participants were asked to select their favorite design scheme from each set and generate illustration materials for the two data analogies. One set of illustration materials was to be generated using the {\systemname} system, while the other set required participants to input keywords into the system or utilize Google Image search to produce images. Participants alternated between the baseline and system conditions, with their condition order randomly assigned (either baseline first and then system or system first and then baseline) to mitigate potential learning effects. All the provided and generated materials will be presented in the supplementary materials.

At the outset, participants were introduced to the fundamental concepts of data analogy design. They were allowed to familiarize themselves with the dataset we had compiled to understand the concept of data analogy. Prior to using the {\systemname} system, they received a brief introduction and demonstration of the system. 
Participants were randomly assigned two sentences with numerical content and were tasked with designing data analogy schemes for them. For each sentence, they were allocated 15 minutes to brainstorm as many design schemes as possible. Subsequently, they were required to generate illustration materials for the selected two proposals independently. We recorded the number of illustration materials ultimately chosen by the participants and the total time spent on generating design materials. Finally, participants were instructed to create the final illustrations using the selected materials on their own. After the experiment concluded, we invited participants to complete the CSI evaluation questionnaire, which measures six dimensions of creativity support~\cite{10.1145/2617588}: \textit{Exploration, Expressiveness, Immersion, Enjoyment, Results Worth Effort}, and \textit{Collaboration}, to assess the usability of our system.

\subsubsection{Result Analysis}
We collected qualitative and quantitative feedback from the participants and summarized them as follows.


\textbf{The amount of analogy ideas.}
Participants generated a significantly larger amount of data analogies than the baseline using {\systemname}.
On average, participants generated five analogies (SD=1.26) using the {\systemname}, whereas the average number of analogies with baseline was 3.873 (SD=1.09). Since this study employed a within-subjects design and involved count data, we conducted a paired-sample Wilcoxon test, revealing that the difference between these means was statistically significant (p < 0.004). In terms of creativity assistance, all the 16 participants unanimously favored {\systemname} for the following reasons: (1) Internet searches for specific keywords relied heavily on individual knowledge and consistently yielded specific keyword data, limiting the exploration of related concepts (P2, P9, P12, and P15); (2) {\systemname} provided diverse creative ideas and inspired creativity (P2, P3, P11, and P12); (3) the interpretation of themes and recommendations of keywords saved time in generating suitable illustration materials (P4 and P6).


\textbf{The time involved in creation. }
{\systemname} enables participants to find illustration materials in significantly less time than the baseline.
On average, participants spent 7.92 minutes generating usable materials with the {\systemname} (SD=2.21), while the average time spent with the baseline was 5.91 minutes (SD=1.91). Since this study employed a within-subjects design and involved count data, we conducted a paired-sample Wilcoxon test, which revealed that the difference between these means was statistically significant (p < 0.007).
When asked which tool was more efficient for finding usable image materials, all 16 participants unanimously agreed that the {\systemname} saved them more time. The two main reasons users mentioned were as follows: (1) The recommended keywords helped them clarify their desired materials. (2) {\systemname} could generate images based on selected keywords, which proved especially effective for sourcing image materials that were not readily available online.

\textbf{System design.} 
In general, users expressed their appreciation for the system's ability to generate analogy designs. As stated by P4 (F, age=22), ``...\textit{When I first looked at the data, I did not have any ideas, but {\systemname} provided many potential analogies and design schemes, some of which proved challenging for me to conceive independently.}'' Moreover, P11 (F, age=22) mentioned that system-generated designs were regarded as inspirations to spark their creativity. 
Additionally, a majority of the users found the system's keyword recommendation and selection feature valuable. For instance, when users feel uncertain about illustration design, the recommended keywords can provide inspiration. As mentioned by P6 (Female, age=20), "\textit{The recommended keywords cover most of the content that I want to convey. Some detailed keywords, such as color temperature and emotions, are more detailed than what I could have thought, and they enhance the final expression of the images.}"


\textbf{The quality of generated results.} 
All participants found the results generated by {\systemname} useful and inspiring. 
Although, despite the two-step correction provided in our pipeline and system, users still reported several inaccuracy occasions due to the limitations of GPT-3.5. For example, P6(F, age=20) mentioned that the ratio of the height of the Himalayas to the height of one step is not exactly 900,000:1. This discrepancy occurred because the height of the Himalayas divided by 900,000 is mistakenly calculated by GPT. Nonetheless, users still found the generated content inspiring and convenient. As in the inaccuracy case P6(F, age=20) encountered, the participant commented, ``\textit{I still think it is helpful to generate the height of the Himalayas as an analogy. I can easily do the simple calculation myself if I want it to be more precise, and sometimes, the accuracy is not a major concern for my targeted audience.}''

\begin{table}
    \vspace{-1mm}
    \begin{tabular}{l | ccc}
    \toprule
    Scale & Avg. Factor Counts (SD) & Avg. Factor Score (SD) & Avg. Weighted Factor Score (SD) \\
    \midrule
    Collaboration & 15.19 (3.15) & 0.63 (0.81) & 9.19 (12.18) \\
    Enjoyment & 17.81 (2.01) & 1.94 (1.18) & 35.16 (23.04) \\
    Exploration & 16.56 (2.53) & 4.19 (0.91) & 69.38 (19.32) \\
    Expressiveness & 15.63 (3.05) & 3.63 (0.96) & 56.19 (17.65) \\
    Immersion & 15.25 (3.53) & 1.5 (0.82) & 23.31 (14.48) \\
    Results Worth Effort & 16.69 (1.89) & 3.13 (1.31) & 51.75 (21.73) \\
    \bottomrule
    \end{tabular}
    \vspace{2mm}
    \caption{CSI Results from a Data Analogy Creation Study Using {\systemname} (N=16). The average CSI score for {\systemname} in this study was 81.63 (SD=9.99).}
    \vspace{-4mm} 
    \label{table:CSIScore}
    \vspace{-4mm}
\end{table}

\textbf{SCI Score of {\systemname}. }
Table~\ref{table:CSIScore} displays the average factor counts, average factor scores, and average weighted factor scores for the six factors of the CSI \cite{10.1145/2617588}. Average factor counts can gauge which factors are most important for the experimental task, with the highest possible count for any specific factor being 5. According to Table~\ref{table:CSIScore}, exploration seems to be the most crucial for participants, while collaboration appears to be less relevant or important to users.
\szd{Each factor has a factor score, which is the total of the agreement statement responses for that factor. The responses are on a 0 to 10 scale, and a higher score indicates that the tool supports that factor better.}
In Table~\ref{table:CSIScore}, we observe that all factors received relatively high ratings, with Exploration, Enjoyment, and Results Worth Effort scoring the highest. However, since exploration is particularly vital for this task, it receives the highest weighted factor score. Weighted factor scores are calculated by multiplying a participant's factor agreement scale score by the factor count to make them more sensitive to the factors most important for the given task.
In the end, we computed the average CSI score using the formula mentioned in \cite{10.1145/2617588} and obtained a score of 81.63 (variance = 9.99), which can be considered a good creativity support score (B-grade). This implies that {\systemname} can provide reasonable creativity support for users engaged in designing data analogies and illustrations but still has room for improvement. Participants were asked to provide explanations for each score, with P9 (F, age=21) giving a lower score for exploration and mentioning the desire to include more historical exploration and exploration records.

\subsubsection{Expert Evaluation.}
To further evaluate the results created by our participants in the user study, we invited two domain experts: a designer with eight years of experience in data visualization (E5) and an academic researcher with over ten years of experience in the interdisciplinary domain of design and technology (E6). We asked the experts to provide their professional opinions and relative ratings on the illustration drafts produced by our participants. Their assessments covered four aspects: whether the analogy design makes it easier to understand the data content, whether the analogy design helps to grasp the scale of the given data, whether the analogy design helps engage the audience with the context, and whether the analogy design is creative.

We asked them to think aloud their opinions and ideas any time they wanted during the interview sessions. We also posed follow-up questions whenever we observed that an interviewee's response was unclear or if we sought deeper insights into the specifics of their daily practices.
\rv{We organized and coded the interview data using the same method described in Section 3.1. After analyzing the interview data from the two experts,} we identified several common traits in the works deemed as ``good analogy designs''.

\underline{\textit{Clarity in expressing data and relationships.}} All the works regarded as good representations of data analogies exhibited a high degree of clarity in the expression of data and relationships. For instance, in Fig.~\ref{fig:showcase}, P6 (F, age=20) compared the distance between the connected plastic bottles with the distance from Earth to the Moon.
 These representations are clear and intuitive in their visual presentations, facilitating immediate understanding.

\underline{\textit{Relevance of chosen analogy objects and themes.}} Another common trait of good data analogies was the relevance of chosen analogical objects and themes. For example, in Fig.~\ref{fig:showcase}, P5 (F, age=22) compared the thickness of a sheet of paper (0.104mm) with the thickness of 900,000 sheets of paper (93.6m) to illustrate a 900,000:1 disparity. Using paper invoked associations with paper money, which emphasized the theme of ``wealth disparity''. P11 (F, age=22) associated plastic bottles with the Earth, a linkage that both experts noted as being related to environmental conservation.

\underline{\textit{Metaphorical information embedded within the theme.}} The inclusion of metaphorical information within the theme was also a common characteristic of good data analogy examples. For instance, in P16's (F, age=20) illustration draft, 7 meters of seawater submerged two stories of a building. Both experts noted that this analogy effectively conveys the gravity of the consequences of global warming. It prompts readers to contemplate the repercussions of people and buildings being inundated, offering them meaningful inspiration.

Despite the shared opinions of the common traits for good analogy designs, we also found the two experts have their own personal preferences regarding analogy designs. E2, the academic researcher, tended to prefer analogy objects that were more surprising and creative, while E1, the industry practitioner, leaned towards more familiar analogy objects. This divergence may stem from the two experts' different backgrounds and working scenarios. For example, P4 (F, age=22) compared the amount of melted ice in Greenland to ``40,000 years of the water flow of Niagara Falls.'' E1 praised its creativity and impact, while E2 expressed concerns about the reader's familiarity with Niagara Falls.

\begin{figure}
    \vspace{-2mm}
    \centering
    \includegraphics[width=1\textwidth]{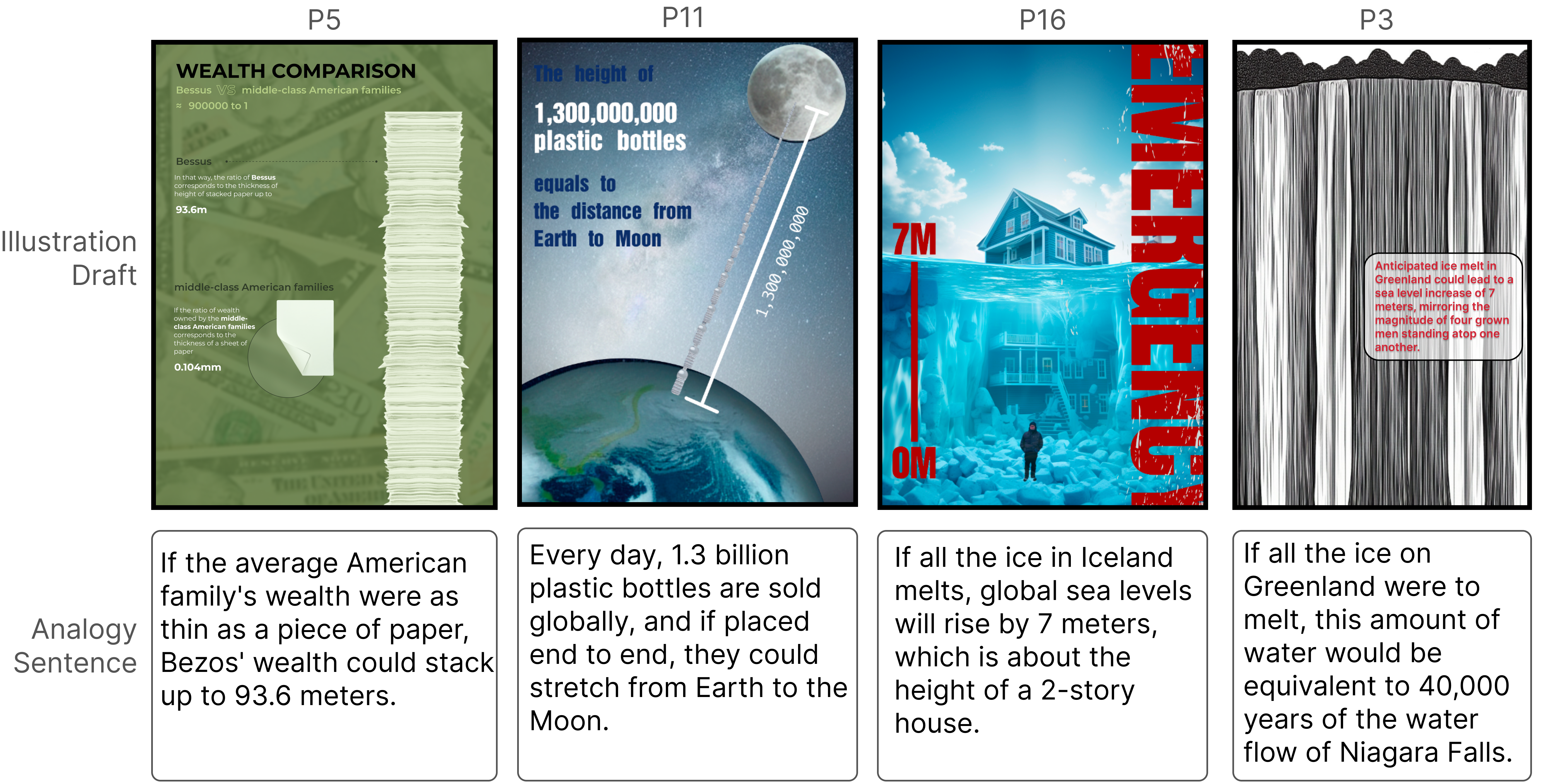}
    \vspace{-6mm}
    \caption{The four analogy design examples created by the participants in User Study I.}
    \vspace{-1mm}
    \label{fig:showcase}
    \vspace{-2mm}
\end{figure}

\rv{\subsection{User Study II: Evaluating Data Comprehension with AnalogyMate}
\begin{figure}
    \vspace{-1mm}
    \centering
    \includegraphics[width=0.95\textwidth]{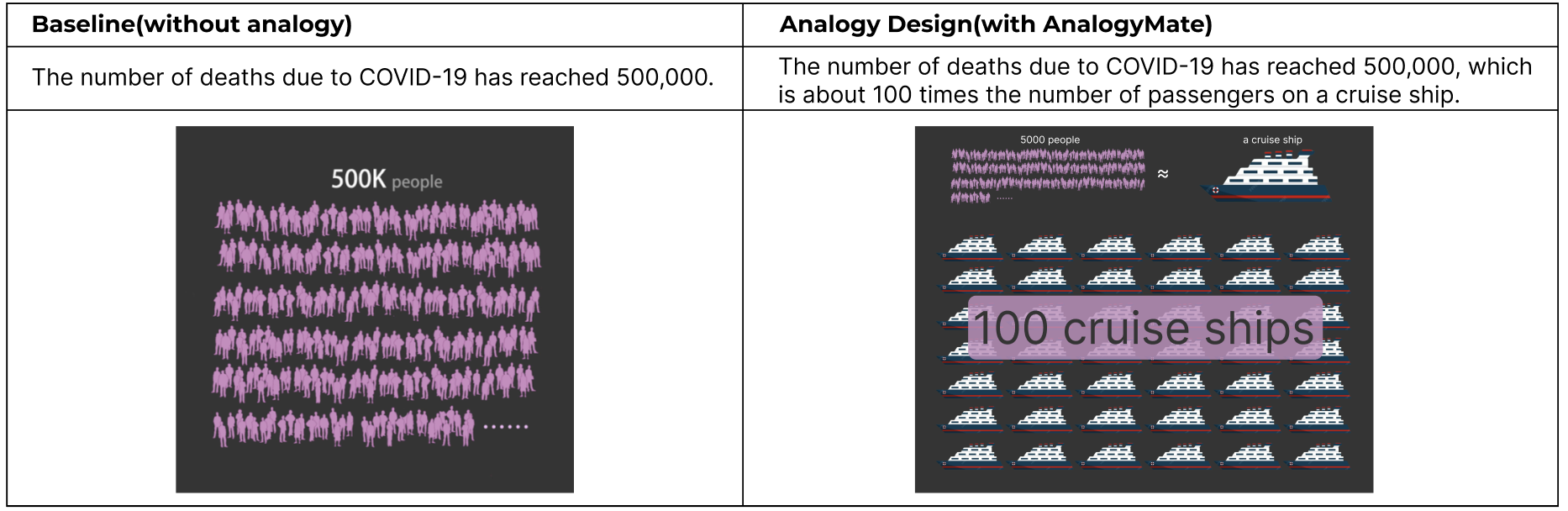}
    \vspace{-1mm}
    \caption{One example case used in User Study II. The baseline without analogy is on the left, while the analogy design created with 
    {\systemname} is on the right.}
    \vspace{-3mm}
    \label{fig:US1}
\end{figure}
To validate the effectiveness of the system-generated data analogies in enhancing data comprehension, we conducted a crowdsourced study, comparing raw data (without analogy) with results generated by {\systemname}.


\subsubsection{Experiment Design}

We adapted the methods used by Sun et al.~\cite{10.1145/3511892}, in which a crowdsourcing task is designed to collect participants' feedback on their understanding of autogenerated data. In general, our study aims to collect participants' responses regarding three perspectives: their understanding of the given data, the extent of measurement of the data scale, and the degree of engagement with the context.


\subsubsection{Participants}

We posted our study form online and recruited participants through the \textit{Prolific} crowdsourcing platform, restricting potential participants to those residing in the US or UK (who speak fluent English) to reduce language barriers. Notifications were provided for participants to acknowledge the study's purpose and the fact that no sensitive information was required. 
We recruited 5 participants for a pilot study and a total of 80 participants for the main experiment.
The experiment was only available to participants who did not enroll in our pilot study. Contributors were rewarded $\$ 1.24$ for an estimated (and actual) completion time of 10 minutes. The time estimation was based on contributions from in-person participants. 

\subsubsection{Procedure and Tasks}
In the user study, eight analogy cases were selected for testing, with each analogy strategy comprising two, covering each topic. Each case presents visualizations of the baseline (raw data, referred to as \textit{infoPoor}) and the data analogy design (\textit{infoRich}). By rearranging them, we designed eight user study forms of 18 pages each. Each stimulus occupies one page (16 pages in total). 
The experiment starts on page 2, with notifications on the first page and validation checks on the last.

On each stimulus page, participants were asked to rate according to the following three statements:
\begin{enumerate}
    \item The statement and graphics make it easier for me to understand the data \textbf{(S1)}.
    \item The statement and graphics help me grasp the scale of the given data \textbf{(S2)}.
    \item The statement and graphics engage me with the context \textbf{(S3)}.
\end{enumerate}
Participants were allowed to revisit previous pages at any time during the experiment. 

Fig.~\ref{fig:US1} presents an example case in our user study, with the \textit{infoPoor} stimuli on the left, \textit{infoRich} stimuli on the right.
The aforementioned three rating statements were listed under every stimulus. 
After finishing all the questions, every participant must complete our validation check by confirming they had done our user study carefully. 
The validation check contains the first stimulus that appears in the given form and one stimulus that does not appear. 


\subsubsection{Result Analysis}
We got valid data from $N=80$ participants ($N=10$ per form). The number was chosen to achieve an available sample size for the t-test. Participants are 46.0\% female, with a mean age of 39 and a standard deviation of 13.4. Most of them are from the UK (66.3\%).

Out of the 113 contributors who agreed to the consent form, 33 did not finish the experiment, resulting in an attrition rate of 29.3\%. Among those, 54.5\% failed the validation check, and 45.4\% quit for unknown reasons. We then manually examined the time of completion and rating sequence to drop out obvious duplicate answers. To further eliminate threats to validity, we compared the standard deviation of rating scores between the \textit{infoRich} and \textit{infoPoor} to make sure participants did not drop out considerably more often in one condition than the other. The difference in standard deviations between the two distributions inforich and infopoor are 0.05 \textbf{(S1)}, 0.01 \textbf{(S2)}, 0.06 \textbf{(S3)}. Thus, we found no evidence that rating scores were distributed differently across two aspects of the same data analogy case.
In other words, participants did not exhibit significant differences in the distribution of ratings when evaluating different aspects of data analogy cases. This contributes to ensuring the stability and consistency of the evaluation results, enhancing the credibility of the research findings.}

\textbf{S1} revolves around the understanding of data, which refers to the ability to interpret, analyze, and make meaningful sense of information presented in a structured or unstructured form. \textbf{S2} focuses on the ``scale of data'', or the level of measurement or the type of data you are dealing with. \textbf{S3} quantifies how well the audience is engaged with the analogy context. We illustrate a t-test to determine whether there are significant differences between infoRich and infoPoor. We set a null hypothesis $H_0$ that the distribution of the two subsets is the same. One-tail t-test demonstrates a result less than $<1\%$. It is safe to conclude that data analogy enhances the audience's understanding of the data, helps them measure the scale of data more precisely, and participants feel engaged in the data context.

\section{Discussion}
In this section, we further discuss the findings of our study. First, we reflect on some of the lessons learned and design reflections during the implementation of {\systemname}. Next, we explore the generality of data analogies and propose potential application domains that could benefit from our tool. Finally, we summarize the limitations of our study and outline directions for future work.

\subsection{Design Reflections}

\textit{Leveraging the strengths of AIGC in data analogy design.} Artificial Intelligence Generated Content (AIGC) offers a significant advantage in providing quick access to extensive knowledge. It can respond to user queries with standard information based on its vast knowledge base. 
Moreover, achieving precise and stable mathematical calculations and understanding using the current version of LLMs and Generative Models remains a challenge. 
Therefore, it is essential to leverage AIGC techniques in specific steps where they excel and complement them with human intervention to bridge this gap effectively. 

\textit{Balancing diverse results with precise control in data analogy generation.} In the process of generating data analogies, we observed the impact of utilizing GPT-3.5 for divergent thinking in analogy creation. 
Allowing GPT-3.5 to generate analogies without restrictions on object types yields diverse and imaginative responses, but when imposing limitations on object types, the generated lists tend to converge, resulting in a slight reduction in overall creativity.
Therefore, when harnessing the power of GPT-3.5 to assist in the design process, it is essential to strike a careful balance between leveraging its divergent thinking capabilities and imposing constraints, a balance that should be determined based on the specific usage scenarios and design requirements of the task at hand.

\subsection{Usage Scenario and Application Domains}
\rv{
Our evaluation results demonstrate the substantial potential of data analogies in conveying intricate information in a more accessible manner, thereby enhancing data comprehension and communication effectiveness. Drawing insights from preliminary study interviews, the collected dataset, and feedback from participants in user studies, we ascertain its broad applicability across diverse domains. Notable application areas include news reporting, where data analogy proves valuable in presenting challenging numerical information in a more understandable form (E3, E4, P11). In popular science and education materials, data analogy serves as a powerful medium for explaining complex concepts to audiences with varying levels of numerical proficiency, making the learning process more engaging (P14). In marketing and advertising, data analogy can elevate the appeal and effectiveness of advertisements through conceptual transformations and visual effects, capturing the interest of potential customers (E2). Furthermore, in cross-industry collaboration and reporting, the use of data analogy facilitates better understanding among participants from different domains, encouraging participation, cooperation, and knowledge transfer in discussions (E3, E4, P11).
}

\subsection{Limitations and Future Work}

This work has several limitations which could provide avenues for future exploration. 

First, due to the scope of our research, we primarily focused on designing data analogies with the assistance of large language models which can only provide general knowledge. Nevertheless, individuals from different regions, age groups, and cultures may possess varying levels of familiarity with certain concepts. Therefore, a future focus is the integration of knowledge bases alongside large language models to tailor analogies to specific audience groups.

Second, the current analogy objects are generated by GPT-3.5, which may introduce occasional errors in size measurements and calculations, despite our implementation of various strategies to minimize such inaccuracies.
\rv{Additionally, there are numerous ethical and responsibility-related issues associated with the generated models that warrant discussion~\cite{xiao2023let}. These issues may result in biases in the generated analogies and illustration materials, as well as an increased risk of copyright infringement. While some research is focused on mitigating these biases~\cite{luccioni2023stable, schramowski2023safe}, further studies are still required. Therefore, fostering critical thinking is crucial for both creators and audiences of AI-generated content.}

Third, in {\systemname}, the unitization strategy may yield results with large numerical values, potentially challenging for users to perceive. 
In future work, our goal is to allow users to choose whether they want to redesign the results for enhanced perceptibility when necessary.

Finally, our future endeavors involve the integration of personal attributes with large language models to facilitate personalized design recommendations. Additionally, we aspire to develop a more cohesive and integrated tool that enables the direct output of well-designed illustrations by the system, providing a more seamless user experience.

\section{Conclusion}
In this work, we introduced {\systemname}, a design support tool that facilitates data analogy creation. Specifically, we first characterized a design space of data analogy based on the collected dataset, previous literature, and preliminary interviews. Then, we proposed an automatic pipeline for data analogy creation and implemented a prototype system following the pipeline. 
Our prototype system has been proven to significantly assist users in generating more creative ideas and reducing the time spent searching for illustration materials. The user study results further demonstrated the effectiveness of the data analogies created with AnalogyMate in enhancing data comprehension and communication.

\begin{acks}
The corresponding author of this paper is Nan Cao. This work was supported in part by the NSFC 62372327, 62002267, 62072338 and NSF Shanghai 23ZR1464700. We would like to thank Xuechen Li for the interface design and our domain experts, including Guanhong Liu, Xian Bai, Yanqiu Wu, Xingyu Lan, Lai Xu, and Yumin Hong for their valuable advice. Additionally, we extend our thanks to the anonymous reviewers for providing constructive feedback.
\end{acks}

\bibliographystyle{ACM-Reference-Format}
\bibliography{main}


\end{document}